\documentclass[reprint,amsmath,amssymb,aps,floatfix,superscriptaddress,longbibliography]{revtex4-2}

\usepackage{graphicx}
\usepackage{dcolumn}
\usepackage{bm}
\usepackage{subcaption}
\usepackage{soul}
\usepackage{comment}
\usepackage[normalem]{ulem}
\usepackage{float}
\usepackage{xr-hyper}
\usepackage{hyperref}
\usepackage[mathlines]{lineno}
\usepackage[nodisplayskipstretch]{setspace}
\usepackage[dvipsnames]{xcolor}

\begin{document}





\title{Division Strategies Determine Growth and Viability of Growing-Dividing Autocatalytic Systems}

\author{Parth Pratim Pandey}
\email{parth.pratimji@gmail.com}
\affiliation{Department of Physics, School of Advanced Engineering,}
\affiliation{Centre for Stochastic Modelling and Simulation,\\ UPES, Dehradun 248 007, India}

\author{Sanjay Jain}
\email{Corresponding Author jain@physics.du.ac.in}
\affiliation{Department of Physics and Astrophysics, University of Delhi, Delhi 110 007, India}
\affiliation{Santa Fe Institute, 1399 Hyde Park Road, Santa Fe, NM 87501, USA}
\affiliation{Department of Physics, Ashoka University, Sonipat, Haryana 131 029, India}


\begin{abstract}
We present a geometric framework to study the growth–division dynamics of cells and protocells, and demonstrate that self-reproduction emerges only when a system’s \textit{growth dynamics} and \textit{division strategy} are mutually compatible. Using several commonly used models (the linear Hinshelwood cycle and nonlinear coarse-grained models of protocells and bacteria), we show that, depending on the chosen division mechanism, the \textit{same} chemical system can exhibit either (i) balanced exponential growth, (ii) balanced nonexponential growth, or (iii) system death (where the system either diverges to infinity or collapses to zero over successive generations). In particular, we show that cellular trajectories in an N-dimensional phase space (where N is the number of distinct chemical species) shuttle between two N-1 dimensional surfaces - a \textit{division surface} and a \textit{birth surface} - and that the relationship between these surfaces and the growth trajectories determines something as fundamental as cellular homeostasis and self-reproduction. Geometrically visualizing bacterial growth and division uncovers, for the first time, the types of division processes that sustain or destroy cellular homeostasis in autocatalytic chemical systems, thereby offering strategies to stabilize or destabilize growing-dividing systems. This reveals that, in addition to autocatalysis of growth, division mechanisms are not passive bystanders but active determinants of a growing-dividing system’s long-term fate. Our work thus provides a framework for further exploration of growing-dividing systems that will aid in the design of self-reproducing synthetic cells.
\end{abstract}

\maketitle

Coarse-grained mathematical models of growing-dividing cells and protocells are of much current interest and have contributed to the understanding of bacterial dynamics and cell-to-cell variability \cite{morgan2004framework, osella2014concerted, amir2014cell, iyer2014scaling, campos2014aconstant, taheri2014cell, harris2016relative, susman2018individuality, pandey2020exponential} as well as protocellular homeostasis and evolution \cite{ganti1975organization, munteanu2006phenotypic, serra2007synchronization, rocheleau2007emergence, mavelli2007stochastic, kamimura2010reproduction, serra2017modelling, hordijk2018population, kahana2023attractor, singh2023multistable}.
Such models involve the nonlinear chemical dynamics of pools of interacting molecular species describing cellular growth, and effective rules for the division process that describe (i) what triggers cell division and (ii) the configuration of daughter cells. 

Such models typically exhibit trajectories for a single cell lineage in which all the variables settle down into a periodic pattern representing self-reproduction of the cell across successive generations. These trajectories represent the spontaneous emergence of `balanced growth' \cite{campbell1957synchronization} wherein none of the intracellular chemical species gets progressively diluted or accumulated across generations. Thus the cell maintains its chemical diversity through repeated rounds of growth and division and avoids a `death by dilution' \cite{luisi2006approaches, serra2017modelling}.
However, there is not much research investigating the conditions in which this homeostasis can be lost. Experimental examples of the loss of balanced growth and cellular homeostasis have been reported \cite{cohen1954studies, beck1962metabolic, schmidt2014loss} where cells have been observed to grow progressively smaller or larger until growth is arrested. Theoretical models of protocells also exist \cite{filisetti2010non, mavelli2013theoretical, serra2019sustainable, serra2017modelling,villani2025protocells}, in which the loss of homeostasis has been shown to arise from specific choices of model parameters that define the kinetic equations governing protocell growth. These interesting examples, together with the intrinsic importance of cellular homeostasis and self-reproduction have motivated us to study the subject more systematically here. In particular, we highlight how the division process constrains growth trajectories and shapes cellular homeostasis - an aspect that has received little attention.\\

\textbf{Growth-Division Dynamics:} In this work, we generalize the physics of growth-division dynamics by developing a geometric framework for understanding these processes. In the coarse-grained models we consider, the state of a cell at any time $t$  is defined by a point $X(t) = (X_1(t),\ldots,X_N(t))$ in an $N$ dimensional phase space $\Gamma$, where $X_i$ is the population of its $i^{\text{th}}$ molecular species, $i=1,\ldots,N$. The \textit{growth-division dynamics} (GDD) tracks a single lineage of growing and dividing cells and has three components:\,(i) an autocatalytic population dynamics wherein the $X_i$ grow with time, (ii) a state-dependent `division control variable' $D(X)$ which triggers division into two cells when it crosses a certain threshold, and (iii) a rule (which we refer to as the `birth map') that defines the state of the tracked daughter at birth. (ii) and (iii) together constitute the `division mechanism' in the models.  

We show that for the same chemical growth dynamics the system can exhibit vastly different behaviors - ranging from exponential balanced growth to nonexponential balanced growth to system death - based solely on the chosen division mechanism. A single cell will be said to exhibit `balanced growth' if the GDD converges to a periodic attractor. In such an attractor the daughter cell at birth is identical from generation to generation; the cell exhibits self-reproduction. If in successive generations the cell volume $V$ or any of the $X_i$ grows and eventually goes to infinity or shrinks to zero and then stays there, we refer to this as `cell death'. `Exponential growth' is a special case of balanced growth wherein (almost) all $X_i$ (except those corresponding to chemicals with very low copy numbers such as the DNA molecule) and $V$ grow exponentially with time between birth and division at the same rate. Exponential growth implies that the ratios of such chemical abundances ($X_i/X_j$) and their concentrations ($X_i/V$) in the cell are constant between birth and division. However, balanced growth as defined above can occur even without exponential growth. In such a case the ratios of chemical abundances and concentrations of the chemicals in a single cell will be periodic but need not be constant in time. Nevertheless, a culture containing a large number of such cells in different phases of their growth-division cycle will exhibit balanced growth in the sense of Campbell \cite{campbell1957synchronization} (constancy of ratios of chemical abundances) to a good approximation when suitable averages are taken across the culture. 


In particular we show, in examples of autocatalytic growth dynamics, that various generic division control variables $D(X)$ with `degree' $>$ 0 robustly lead to spontaneous emergence of balanced growth. ($D(X)$ has degree $\alpha$ if $D(\rho X) = \rho^{\alpha}D(X)$ for all $X$ and for all $\rho >0$.) Examples of such division variables are chemical abundances, volume, $X_i^{\alpha}$ with $\alpha > 0$, surface area, `reduced surface' \cite{mavelli2007stochastic, mavelli2013theoretical}, etc. The growth may be exponential or otherwise depending upon the birth map. Whereas, if the division variable is intensive (degree zero) -- e.g., a chemical concentration or a ratio of chemical populations $X_i/X_j)$ -- then the growing-dividing system can fail to attain a state of self-reproduction and ultimately die. We also identify strategies (birth maps) that can make such division variables work. We hence show that balanced growth, while generic and robust, cannot be taken for granted for autocatalytic systems and requires the mutual compatibility of the growth dynamics and the division mechanism.\par

The three components of GDD are the following: \\ \textbf{C1 - Growth Dynamics}: This is a rule that determines how the chemical populations $X$ within a single cell change with time between birth and division, starting from a given initial condition at birth. In the present work we restrict our examples to the deterministic dynamics specified by differential equations of the form $dX_i/dt = f_i(X), \quad i=1,2,...,N$. The functions $f_i$ depend upon the populations $X$ which are time dependent, as well as the parameters characterizing the organism and the medium which are assumed constant. $f_i$ contain all information about the possible chemical reactions inside the cell including transport reactions involving exchange of chemicals with the environment. We are primarily interested in the case where the set of reactions is autocatalytic, giving rise to functions $f_i$ such that the $X_i$ typically grow with time after initial transients.\\
\textbf{C2 - Division Control Variable} $D$: As the state $X$ changes with time according to \textit{C1}, we assume that the cell effectively makes a decision to divide whenever the state function $D=D(X)$ crosses a threshold value $d$. 
Examples of some simple division control variables commonly used to model growing-dividing systems are (i) cell volume, $D(X)=V(X) =\sum_iX_i$ (often the total number of bulk molecules in the cell is taken as a surrogate for cell volume) \cite{furusawa2003zipf,morgan2004framework,munteanu2007generic,kamimura2010reproduction,sharma2018modeling,trickovic2022resource,villani2025protocells}, (ii) abundance $X_i$ of a particular intracellular chemical species \cite{serra2007synchronization, carletti2008sufficient,harris2016relative,susman2018individuality,pandey2020exponential,serbanescu2020nutrient,bertaux2020bacterial,panlilio2021threshold}, (iii) total surface area of the cell defined in terms of the population of a membrane-forming chemical species \cite{ganti1975organization,munteanu2006phenotypic,rocheleau2007emergence}, and (iv) reduced surface (a variable characterizing vesicle membrane instability) of the system \cite{mavelli2007stochastic, mavelli2013theoretical}. 
As a convention, cell division is triggered when $D(X)$ reaches the value $d$ from below.
Once triggered, division results in the formation of two daughter cells. \\
\textbf{C3 - Birth-map}: 
The birth map $B:\Gamma \rightarrow \Gamma$ defines the state $B(X)$ of a newborn daughter cell, given the state $X$ of the mother at division. The most commonly considered birth map (which we refer to here as the \textit{standard birth map} is $B_i(X) = X_i/2$ for all $i$. This corresponds to the case of a division process that produces two identical daughters each getting half the population of every chemical in the mother cell at division. This has been used in a number of models (\cite{furusawa2003zipf,munteanu2006phenotypic,mavelli2007stochastic,filisetti2010non,mavelli2013theoretical,sharma2018modeling,serra2019sustainable,trickovic2022resource,serra2007synchronization,susman2018individuality}). 
An example of a \textit{non-standard birth map} considered in various models \cite{harris2016relative,si2019mechanistic, pandey2020exponential,serbanescu2020nutrient,panlilio2021threshold} is one where one of the chemical populations $X_n$ is not halved in the daughter cells but is reset to a fixed value $r\geq 0$, i.e., $B_i(X) = X_i/2$ for all $i \neq n$ and $B_n(X) = r$. This is biologically motivated by the observed reset of certain molecular species between the triggering and the completion of the division process, such as the FtsZ protein or peptidoglycan precursor molecules at division \cite{harris2016relative,si2019mechanistic,den2017divisome} or the active DnaA protein bound to the site of the origin of DNA replication at the initiation of replication \cite{katayama2017dnaa, fu2023bacterial}. This is just one example of a non-standard birth map and is the one majorly considered in this work. A `mass-preserving' variant of this non-standard birth map is considered in the Supporting Material. In the present work we consider both daughters to be identical, hence they have the same birth map.

The above three components together form the growth-division dynamics (GDD) which constitutes the class of dynamical systems defined by the two statements:\\
\begin{small}
(${\cal A}_1: Growth$) If $ D(X) < d$, follow \textit{C1}. \\
(${\cal A}_2: Division$) If $D(X) \geq d$, $X$ is replaced by $X' = B(X)$.
\end{small}
\\
As defined above GDD typically consists of a period of growth while $D(X)< d$, followed by instantaneous division when $D(X)$ becomes equal to the `division threshold' $d$. The configuration $X$ at which the cell divides is referred to as the `mother-at-division'; $X'=B(X)$ then defines the configuration of the tracked `daughter-at-birth' in the next generation. If $D(X') < d$ (and this is typical of cases where balanced growth arises), ${\cal A}_1$ applies after birth and the cell grows again, resulting in repeated rounds of growth and division. Each round of dynamics between two successive divisions (or two successive births) corresponds to a generation.

Ideally, the division variable and birth map should emerge from a single dynamical process governing the time evolution of the cell's state from birth to the completion of division. However, such models would be more complicated.
The GDD formalized here is an approximate phenomenological description of the dynamics in terms of `effective' constructs that has the virtue of simplicity, being formulated in terms of only time dependent populations. Further, for simplicity, here we assume that there is no time delay between `initiation' of division (which occurs when $D(X)$ becomes equal to $d$) and the actual production of daughter cells.

\textbf{Growth-division steady-state}: Of particular interest is the `\textit{growth-division steady-state}' (GDSS). This is a limit cycle of GDD, a time-dependent periodic trajectory (with period $\tau$) in which the system repeatedly pursues the same path of growth and division in successive generations. In a GDSS by definition a cell exhibits balanced growth. 
Thus, if a GDSS is a stable attractor of the dynamics with a large basin of attraction, balanced growth and self-reproduction are robust outcomes of the dynamics. $\tau$ is referred to as the interdivision time or the duration of each generation in the GDSS.\\

\noindent\textbf{RESULTS:}
\textbf{Geometry of the growth-division dynamics}: It is useful to introduce the {\it division surface} $S_D$, the $N-1$ dimensional hypersurface in $\Gamma$, defined as the set of points $X$ satisfying the constraint $D(X) = d$. $S_D$ partitions the physical phase space (the non-negative orthant of $\Gamma$) into a `growth region' $D(X) <d$ and a `division region' $D(X) \geq d$. The $N-1$ dimensional {\it birth surface} $S_B$ is defined as the image of $S_D$ under the birth map $X \rightarrow B(X)$. A cell that starts in the growth region follows a continuous trajectory according to \textit{C1} until it hits a point $X$ on $S_D$. Then the trajectory instantaneously (and  discontinuously) jumps to the point $X'=B(X)$ on $S_B$. If $D(X') <d$ it again moves continuously till it hits $S_D$, and so on. A GDSS defines a curve from $S_B$ to $S_D$ that is traced again and again. A cell that starts in the interior of the division region will immediately divide (according to ${\cal A}_2$), and will suffer repeated divisions until it either enters the growth region, or dies. In this work a cell will be said to die if its trajectory gets confined to one of the boundaries of the physical phase space (at least one of its chemical populations $X_i$ is lost for ever) or if one of the $X_i$ becomes infinite.

The GDD framework has also been used in models with stochasticity present in the growth dynamics, birth map or division control 
(see, e.g.,  \cite{villani2014growth,ghusinga2016mechanistic,pandey2020exponential}). For concreteness and simplicity in this work we restrict ourselves to the deterministic version of GDD as defined above. The geometric considerations discussed here are also useful in thinking about the stochastic versions.

\begin{center}
\begin{figure*}[!htbp]
\centering
\captionsetup{font={footnotesize}}
\includegraphics[width=17.5cm,height=8cm]{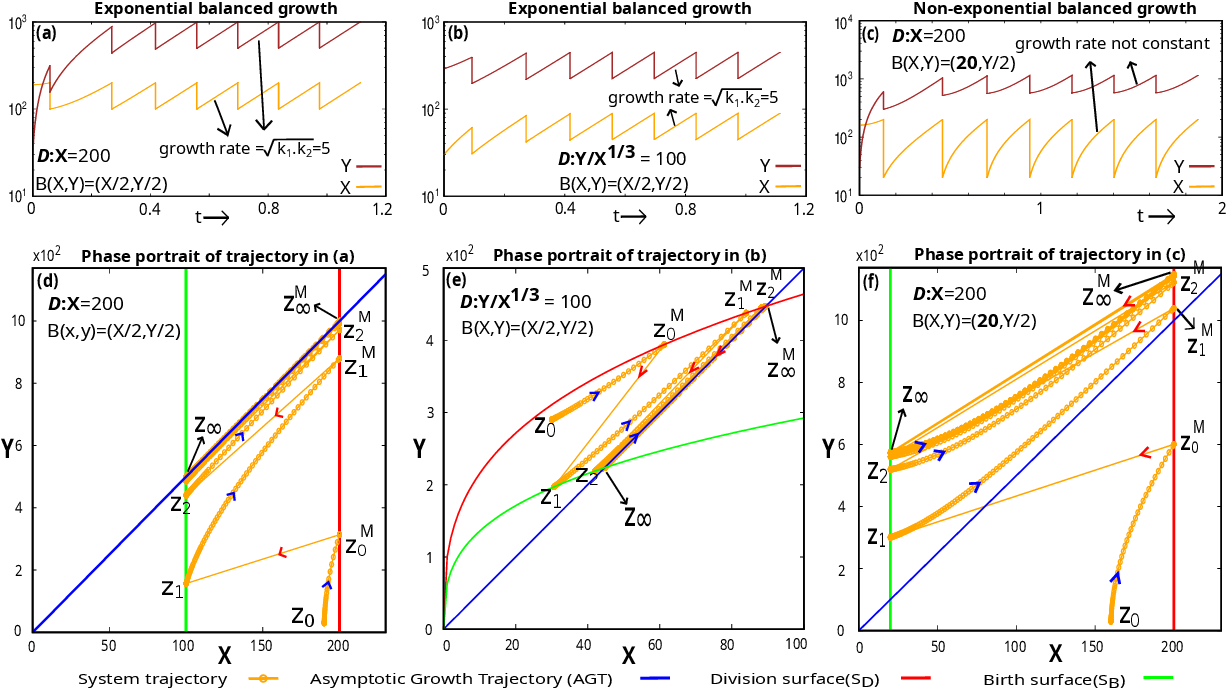}
\caption{Exponential or non-exponential balanced growth of the growing-dividing Hinshelwood 2-cycle with different division mechanisms. Parameter values: $k_1=1$, $k_2=25$. (a), (b) and (c) show trajectories as plots of $X$ and $Y$ versus $t$. (a) Stable \textit{exponential} GDSS reached with division control variable $D=X$, division threshold $d=200$ and standard birth map $B(X,Y) = (X/2, Y/2)$. IC: $(X,Y) = (190,30)$. (b) Stable \textit{exponential} GDSS reached with $D=Y/X^{1/3}$, $d=100$ and standard birth map. IC: (30,290). (c) Stable \textit{non-exponential} GDSS reached with $D=X$, $d=200$ and a non-standard birth map $B(X,Y)=(20,Y/2)$. IC: (160,30). 
Notice that the growth trajectories become exponential (straight line segments in a semi-log plot) in (a) and (b) at the steady-state with the expected rate $\lambda = \sqrt{k_1k_2}=5$. But in (c) the steady-state trajectories are \textit{not} exponential. (d), (e) and (f) plot the same trajectories in phase space for (a), (b) and (c) respectively. Orange lines with circles and blue arrows are growth trajectories. Orange lines with red arrows are instantaneous jumps at division. The division surface, $S_D$, is the red curve $D(X,Y)$=$d$. The birth surface, $S_B$ is the green curve. The blue line is the asymptotic growth trajectory, AGT.
The growth-division trajectory starts at ${\bf z}_0$ and then passes through the sequence of points ${\bf z}^M_0$, ${\bf z}_1$, ${\bf z}^M_1$, ${\bf z}_2$,... which alternately lie on $S_D$ and $S_B$. The trajectory converges to the limit cycle ${\bf z}_\infty$, ${\bf z}^M_\infty$, ${\bf z}_\infty$, which is the GDSS. Note that for the standard birth map the GDSS lies on a segment of the AGT but for the non-standard birth map it does not.
}
\label{fig:xy-autocatalytic-1}
\end{figure*}
\end{center}
\vspace{-9mm}

We now show, using simple autocatalytic models, that the emergence of self-reproduction is contingent upon the mutual compatibility between the three components, growth-dynamics (\textit{C1}), division control variable (\textit{C2}), and the birth map (\textit{C3}).\par
\textbf{GDD on the Hinshelwood 2-cycle}: Consider the $N=2$ system whose growth dynamics (\textit{C1}) is given by:
\begin{equation} \label{eq:xy-ACS}
dX/dt = k_1Y,  \quad dY/dt = k_2X,
\end{equation}
where $X$ and $Y$ are the two chemical abundances and $k_i$ are positive constants. This system has a general solution $X(t) = \sqrt{k_1}(a_1e^{\lambda t} + a_2e^{-\lambda t})$, $Y(t) = \sqrt{k_2}(a_1e^{\lambda t} - a_2e^{-\lambda t})$, where $\lambda = \sqrt{k_1k_2}$ and $a_1$ and $a_2$ depend on initial conditions (ICs). If the system is allowed to grow indefinitely (uninterrupted growth without division) then for generic ICs the system will asymptotically converge to the line $Y = m_{A} X$ with slope $m_A = \sqrt{k_2/k_1}$, which we call the \textit{asymptotic growth trajectory} (AGT) of the system. On the AGT, both the abundances will grow \textit{exponentially} with the rate $\lambda$, which is the larger of the two eigenvalues $\pm \lambda$ of the interaction matrix (see Supporting Material (SM) section S1 for details \cite{SM}). In other words, for generic ICs, `growth' will always push the system towards the AGT.

\textbf{Exponential balanced growth}: Consider the Hinshelwood 2-cycle (Eq. \eqref{eq:xy-ACS}) undergoing GDD with an abundance division variable, $D(X,Y)=X$, division threshold $d$ and a standard birth map. Fig. \ref{fig:xy-autocatalytic-1}a shows the growing-dividing 2-cycle reaching an exponential GDSS where both the chemical abundances grow exponentially with the same rate ($=\lambda$). Fig. \ref{fig:xy-autocatalytic-1}d shows the trajectory in the XY plane. 
The \textit{growth region} corresponds to $D < d$ ($X<200$ in this case). When a trajectory that starts in the growth region at $t = 0$ at the point $z_0 = (X_0 , Y_0)$ reaches $S_D$, the cell divides (applying ${\cal A}_2$). We denote the point of intersection of the trajectory and $S_D$ as $z^M_0$ (`$M$' for `mother'). The newborn daughter then starts at $z_1 = z^M_0/2$ which lies on $S_B$. Since $z_1$ is in the growth region ${\cal A}_1$ applies and the growth dynamics Eq. \eqref{eq:xy-ACS} leads to $z^M_1$ on $S_D$. Then division leads to $z_2$ on $S_B$, and so on. It is geometrically self-evident that the GDD trajectory converges to a segment of the AGT, where at birth the system is always at the point $z_{\infty}$ (where the AGT intersects $S_B$) and at division always at $z^M_{\infty}$ (where the AGT intersects $S_D$). This is the GDSS, a stable limit cycle of GDD. Fig. \ref{fig:xy-autocatalytic-1}b (and its phase space diagram Fig. \ref{fig:xy-autocatalytic-1}e) shows a similar behaviour for another non-intensive division variable, $D = Y/X^{1/3}$. Similar behaviour is observed for $D = X+Y$ (see Fig. S1 in SM). In fact all $D$ variables of type $D = X, Y, X + Y$ or $D = X^{\alpha_1}Y^{\alpha_2}$ ($\alpha_1+\alpha_2 > 0$) will lead to a stable exponential GDSS under the standard birth map.

In SM section S2 \cite{SM} we prove a general result that for the Hinshelwood 2-cycle with the standard birth map, whenever the $D$ variable is such that its corresponding $S_D$ intersects the AGT transversally (not tangentially) once at some nonzero finite point on the plane, every trajectory starting in the growth region $D<d$ will converge to a stable GDSS lying on the AGT (i.e., balanced exponential growth is a robust outcome of the dynamics). This is true for all the $D$ variables mentioned above. A possible generalization of this result for higher dimensional linear dynamical systems is mentioned in SM section S3 \cite{SM}.

\textbf{Non-exponential balanced growth}: 
Fig. \ref{fig:xy-autocatalytic-1}c shows the 2-cycle reaching a non-exponential GDSS with $D=X$, $d=200$ and a non-standard birth map $B(X,Y)=(r,Y/2$), i.e., at division the abundance of $X$ is \textit{reset} to a predefined value (here $r=20$). $X$ starts from $r=20$ and ends at $d=200$ in all the generations (except the initial one where it can start from anywhere in the growth region $X < 200$). Fig. \ref{fig:xy-autocatalytic-1}f shows the phase-space trajectory for this case. SM section S4 gives the analytic formula for the location of $z_\infty$ and the limit cycle trajectory for this case and a proof that this GDSS is stable. It further shows that a `mass conserving' variant of the non-standand birth map also leads to a stable non-exponential balanced growth. The reason for the non-exponential trajectories is the choice of a non-standard birth map. This can be understood geometrically in Fig. \ref{fig:xy-autocatalytic-1}f where it can be seen that the GDSS is \textit{not} on the AGT anymore. Note that the general solution to \eqref{eq:xy-ACS} is a mixture of two exponential functions of $t$. Only when the IC lies on the AGT, i.e., when $a_2=0$, do we get a pure exponential trajectory. 
As shown in Fig. S2 (in SM), farther away the reset value $r$ is from $d/2$, the farther away GDSS is from the AGT, and the greater the deviation from exponentiality in the GDSS. The example in Figs. \ref{fig:xy-autocatalytic-1}c and \ref{fig:xy-autocatalytic-1}f shows that even though the growth dynamics \eqref{eq:xy-ACS} pushes the trajectory towards the AGT, a non-standard birth map results in the stable attractor of the GDD not being on the AGT. In the case of the standard birth map (illustrated in Figs. \ref{fig:xy-autocatalytic-1}a,b,d and e) the GDD attractor lies on a segment of the AGT resulting in a GDSS with exponential growth. The underlying geometric reason is that the AGT is invariant under the standard birth map (a point on the AGT is mapped to another point on the AGT), while it is not invariant under the non-standard birth map. We remark that partitioning stochasticity at division also throws the trajectory further away from the AGT and causes fluctuations of the observed growth rate of cells \cite{pandey2020exponential} for the same geometric reason.
\begin{figure}[!t]
\centering
\captionsetup{font={footnotesize}}
\includegraphics[width=8.5cm, height=8cm]{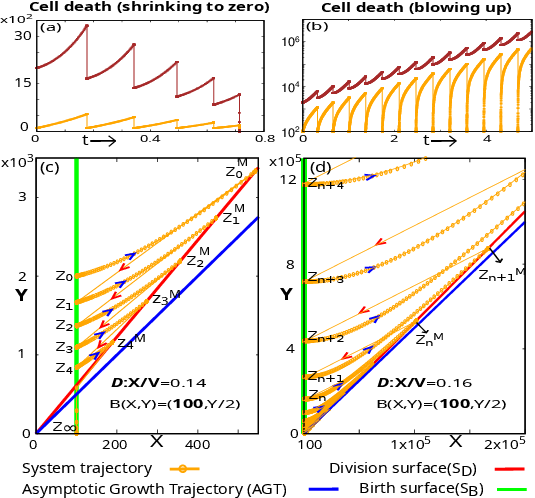}
\caption{Division processes when the growing-dividing Hinshelwood cycle (Eq. \eqref{eq:xy-ACS}) \textit{fails} to reach a self-reproducing state. In both figures the division process is defined by the intensive division variable $D=X/V$ (where $V=X+Y$), division threshold $d$, and the non-standard birth map $B(X,Y) = (100,Y/2)$. $S_D$ is a straight line (red) passing through the origin. Conventions are the same as in Fig. \ref{fig:xy-autocatalytic-1}. IC: (X,Y) = (100,2000) for both plots. $k_1=1$, $k_2=25$. (a) $d=0.14$. The system progressively becomes smaller after every division, and asymptotically shrinks to zero. (b) $d=0.16$.  For the IC considered, the system progressively becomes larger after every division and asymptotically blows up to infinity. In (a) $m_D > m_{max}$. ($m_D \equiv (1-d)/d = 6.14$, $m_{max} \equiv (2/\sqrt{3})m_{A} = 5.77$.) In (b) $m_{A} < m_D < m_{max}$. ($m_{A} =\sqrt{k_2/k_1}= 5$, $m_D = 5.25$, $m_{max} = 5.77$.)}
\label{fig:xy-autocatalytic-2}
\end{figure}

\textbf{Death by division with intensive division variables:}
An important class of division variables to consider are chemical concentrations $X/V$ or $Y/V$. It has been noted \cite{fu2023bacterial} that while in eukaryotic cells concentration thresholds are important checkpoints, in bacteria chemical concentration thresholds are not suitable triggers because concentrations seem to be constant across the cell cycle. 
When the cell volume is a degree one function of the populations (for example $V = X+Y$), the concentrations are intensive variables. Consider the case $D= X/(X+Y)$ with threshold $d$. Note that by definition $d<1$. The division surface $S_D$, defined by $X/(X+Y) = d$, is the straight line through the origin with positive slope $m_D = (1-d)/d$ (see Fig. \ref{fig:xy-autocatalytic-2}). The growth region (where $D <d$) is the region above this line in the positive $XY$ quadrant. 

The intensivity of $D$ implies that the value of $D(X,Y)$ is unchanged when $(X,Y)$ is transformed under the standard birth map. Physically, the standard birth map halves both $X$ and $V$ and therefore does not change the concentration of $X$. Geometrically, the standard birth map leaves $S_D$ invariant ($S_B=S_D$) and hence is unable to transport the daughter cell to the growth region. Therefore, for an intensive $D$ variable to work we need to use a non-standard birth map. Notice that the above argument applies in any dimension $N$, and for any growth dynamics.

To implement the non-standard birth map in our 2-cycle, one can either reset $X$ or $Y$. We first show that with $D=X/V$ and the non-standard birth map $B(X,Y)=(r,Y/2)$ (i.e., resetting $X$), we do not get a stable limit cycle; we get cell death. 
Depending upon the values of $k_1$, $k_2$ and $d$, three generic cases arise: (a) If $m_D > m_{max} \equiv (2/\sqrt{3})m_{A}$ then no periodic orbit exists and all growing-dividing trajectories starting in the growth region shrink to $Y=0$ (for an example see Fig. \ref{fig:xy-autocatalytic-2}a). (b) If $m_{A} < m_D < m_{max}$ then there exists a periodic orbit starting from a point $(r, Y^*)$ on $S_B$, but is unstable (for proofs see SM section S5 \cite{SM}). Trajectories starting from $S_B$ with $Y > Y^*$ eventually blow up to $Y=\infty$, as shown in Fig. \ref{fig:xy-autocatalytic-2}b. Those starting from $S_B$ with $Y<Y^*$ shrink to $Y=0$ (Fig. S3). (c) If $m_D < m_{A}$ then all trajectories starting in the growth region asymptotically approach the AGT and go to infinity without any division (Fig. S4).

\textbf{Making intensive variables work:}
We find that the other non-standard birth map $B(X,Y) = (X/2,r)$, i.e., resetting $Y$ instead of $X$, gives a stable GDSS with the division variable $D = X/V$. Equivalently, the birth map $B(X,Y) = (r, Y/2)$ gives a stable GDSS for the division variable $D= Y/V$. This is shown in Fig. S5. In SM section S6 a proof of stability of this GDSS is given. Note that in this case the variable that is reset at division is different from the one whose concentration triggers division.

\textbf{GDD for a nonlinear protocell  model:} Generalizing our results to nonlinear autocatalytic models in higher dimensions, in SM section S7 we show that a protocell model based on well-stirred, non-linear mass-action chemical kinetics exhibits the same behaviour as the linear Hinshelwoood 2-cycle, when subjected to GDD. The protocell model has three species, a precursor molecule $P$, a lipid molecule $L$ and a catalyst $C$, and is similar to a coarse grained model for bacterial cells \cite{pandey2016analytic, pandey2020exponential} that reproduces several experimentally observed phenomena in \textit{Escherichia coli}.
While the nonlinearity of the protocell model prevents us from giving analytic proofs that have been provided for the linear 2-cycle, we give numerical evidence for the same kind of behaviour as discussed above for the 2-cycle.\\

\noindent\textbf{DISCUSSION:}
Our examples, both linear and nonlinear, are limited to autocatalytic systems whose growth dynamics have an AGT that is a straight line passing through the origin. The latter is a consequence of the homogeneous degree-one character of the growth functions $f_i(X)$ \cite{pandey2020exponential}, which, in turn, is a consequence of the system volume $V$ being a homogeneous degree-one function of the chemical populations. While this class of systems is relevant for biology \cite{pandey2020exponential,lin2020origin, Salman2025Emergent}, it would be interesting to consider other growth dynamics as well.


The models considered here assume that the cell `senses' the value of the division control variable $D$ and division is triggered when $D$ reaches a threshold value $d$. For a cell, a concentration variable is chemically easier to sense (e.g., by local receptors) than the absolute population of a chemical (if the chemical is distributed across the cell). However, concentration is an intensive variable, and, as seen above, does not naturally lead to balanced growth when used as a division variable in the growth-division dynamics discussed here, except with somewhat unusual birth maps. In this work we have found a class of birth maps that do produce balanced growth even with concentration type division variables. This observation may be useful in constructing synthetic cells with concentration as the division control variable.\\

\noindent\textbf{ACKNOWLEDGMENT:} PPP acknowledges the Maharishi Kanad Postdoctoral Fellowship from Delhi School of Public Health at Institution of Eminence, University of Delhi, Delhi. SJ acknowledges the International Centre for Theoretical Sciences - TIFR, Bengaluru, where part of this work was done.

\nocite{*}

\bibliographystyle{unsrtnat}
\bibliography{references}

\begin{thebibliography}{46}
\bibitem[SM()]{SM}
See Supporting Material which has proofs, additional details and figures for
  the growing-dividing Hinshelwood 2-cycle and the nonlinear protocell model.
\end{thebibliography}

\end{document}